\documentclass[conference]{IEEEtran}
\IEEEoverridecommandlockouts
\usepackage{cite}
\usepackage{amsmath,amssymb,amsfonts}
\usepackage{algorithmic}
\usepackage{graphicx}
\usepackage{textcomp}
\usepackage{multirow}
\usepackage{array}
\usepackage{makecell}
\usepackage{xcolor}
\usepackage{subcaption}
\def\BibTeX{{\rm B\kern-.05em{\sc i\kern-.025em b}\kern-.08em
    T\kern-.1667em\lower.7ex\hbox{E}\kern-.125emX}}
\begin{document}

\title{Experimental Analysis and Modeling of Penetration Loss for Building Materials in FR1 and FR3 bands \\
}
\author{
Enrui Liu\textsuperscript{*}, 
Pan Tang\textsuperscript{*}, 
Tao Jiang\textsuperscript{+} and 
Jianhua Zhang\textsuperscript{*}\\
\textsuperscript{*}
Beijing University of Posts and Telecommunications, Beijing, China\\
\textsuperscript{+}China Mobile Research Institute, Beijing, China\\
\textsuperscript{*}Email: \{liuenrui, tangpan27, jhzhang\}@bupt.edu.cn\\
\textsuperscript{+}Email: jiangtao@chinamobile.com
}

\maketitle

\begin{abstract}
 This study focuses on analysis and modeling of the penetration loss of typical building materials in the FR1 (450 MHz–6 GHz) and FR3 (7–24 GHz) bands based on experimental measurements. Firstly, we measure the penetration loss characteristics of four different typical building materials from 4 to 16 GHz, including wood, glass, foam and concrete, by using a penetration loss measurement platform based on the vector network analyzer (VNA). Next, we analyze the frequency dependence and thickness dependence of penetration loss. Finally, the linear model is applied to fit the curve of the measured penetration loss, and new model parameters for the penetration loss of different building materials are given, which are compared with that in the third generation partnership project (3GPP) technical report (TR) 38.901. The analysis results and new model parameters may provides insight into understanding propagation characteristics in FR1 and FR3 bands and 3GPP channel model standardisation.

\end{abstract}


\section{Introduction}
The FR1 (Frequency Range 1) and FR3 (Frequency Range 3) frequency bands are becoming increasingly important due to the growing demand for spectrum resources and the congestion in lower frequencies. These bands offer valuable additional bandwidth, which is crucial for supporting the expanding needs of modern communication systems\cite{1}. With their ability to provide a balance between coverage and capacity, especially in high-density urban areas, FR1 and FR3 are expected to complement other frequency ranges like mmWave\cite{bib1}, \cite{bib5}.

Wall penetration loss is the difference between the median of the location variability of the signal level on one side of a wall, and the signal level on the opposite side of the wall at the same height above ground\cite{bib6}. It is a critical parameter for designing and deploying wireless communication systems, as it affects the coverage, capacity, and quality of service\cite{bib7}, \cite{bib8}. Given that penetration loss is frequency-dependent and varies significantly across different frequencies, particularly in the FR1 and FR3 bands where frequencies range from 450 MHz to 24 GHz with a nearly 48-fold increase, it is essential to conduct an in-depth study of penetration loss across these two bands. 

Recently, several measurement campaigns have been carried out to study the penetration loss in FR1, FR3, and mmWave frequency bands and typed building materials. In \cite{bib9} and \cite{bib10}, the author presents measurements of penetration losses for common construction materials such as wood, window, brick wall, polystyrene and wooden door at 28 GHz, 38 GHz, and 26.5 to 40 GHz, respectively. In \cite{bib11}, the author presents reflection coefficients and penetration losses for common building materials such as glass and concrete wall at 28 GHz for the design and deployment of future millimeter-wave mobile communication networks. In \cite{bib12}, the author investigates the impact of window penetration loss (WinPL) on the frequency dependence of building entry loss (BEL) from 3.5 to 24 GHz. The WinPL characteristics of an ideal double-glazed glass and an actual double-glazed window are simulated and measured on-site, respectively. In \cite{bib13}, the author presents the results of a window penetration loss (WinPL) and building entry loss (BEL) of a traditional office building from 3.5 to 24 GHz. In \cite{bib14}, the author presents results from extensive material penetration loss measurements in FR1 and FR3 bands for ten common construction materials found inside buildings and on building perimeters. Generally, research on penetration loss in the FR1 and FR3 bands remains limited, with existing studies often focusing on few specific frequency points (such as 6.75 GHz and 16.95 GHz) and single materials (e.g., glass)\cite{bib15}. This paper conducts multi-frequency point measurements and develops new model parameters based on detailed experimental data.

In this paper, we introduce the measurements of different materials penetration loss in FR1 and FR3 bands. The main contributions of this paper are as follows:
\begin{enumerate} 
 \setlength{\itemsep}{-2ex}  
 \setlength{\parskip}{0ex} 
 \setlength{\parsep}{0ex}
\item The penetration loss of four different materials (wood, glass, concrete, and foam) was measured. These materials are typical building materials, and their penetration loss characteristics are of particular interest to 3GPP TR 38.901.\hfil\break
\item The variations in penetration loss with respect to frequency, material type, and material thickness are thoroughly investigated and analyzed. \hfil\break
\item New model parameters for the penetration loss of different building materials in the FR1 and FR3 bands were obtained. A comparison with 3GPP TR 38.901\cite{bib16} reveals that the existing standards generally overestimate the penetration loss values.
\end{enumerate}

The rest of the paper is organized as follows: Section II introduces penetration loss measurement platform, configurations, and measurement methods. Section III presents the analysis and modeling of the measured penetration loss. Finally, Section IV concludes the paper. 

\section{Measurement Platform and Measurement Method}
Channel measurement is essential in telecommunications as it provides critical data on signal propagation, which is fundamental for designing and optimizing communication systems. The setup of the measurement apparatus, the test environment, and the detailed procedures are introduced in the subsequent section.

\subsection{Measurement Platform and Configuration}
\begin{figure}[!ht]
\centering
\subfloat[The main view of the measurement platform.]{
		\includegraphics[scale=0.09]{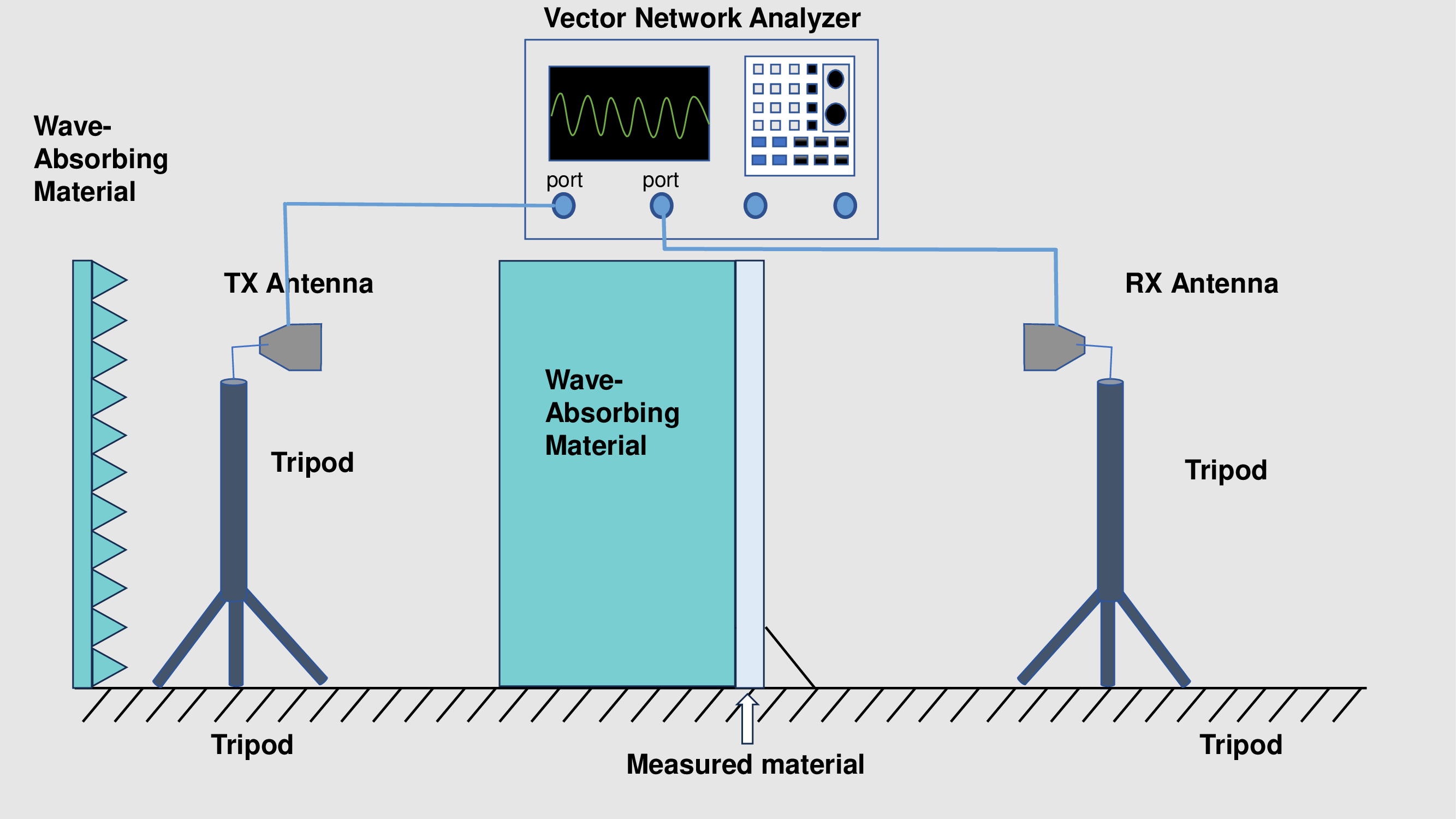}}
\\
\subfloat[The top view of the measurement platform.]{
		\includegraphics[scale=0.09]{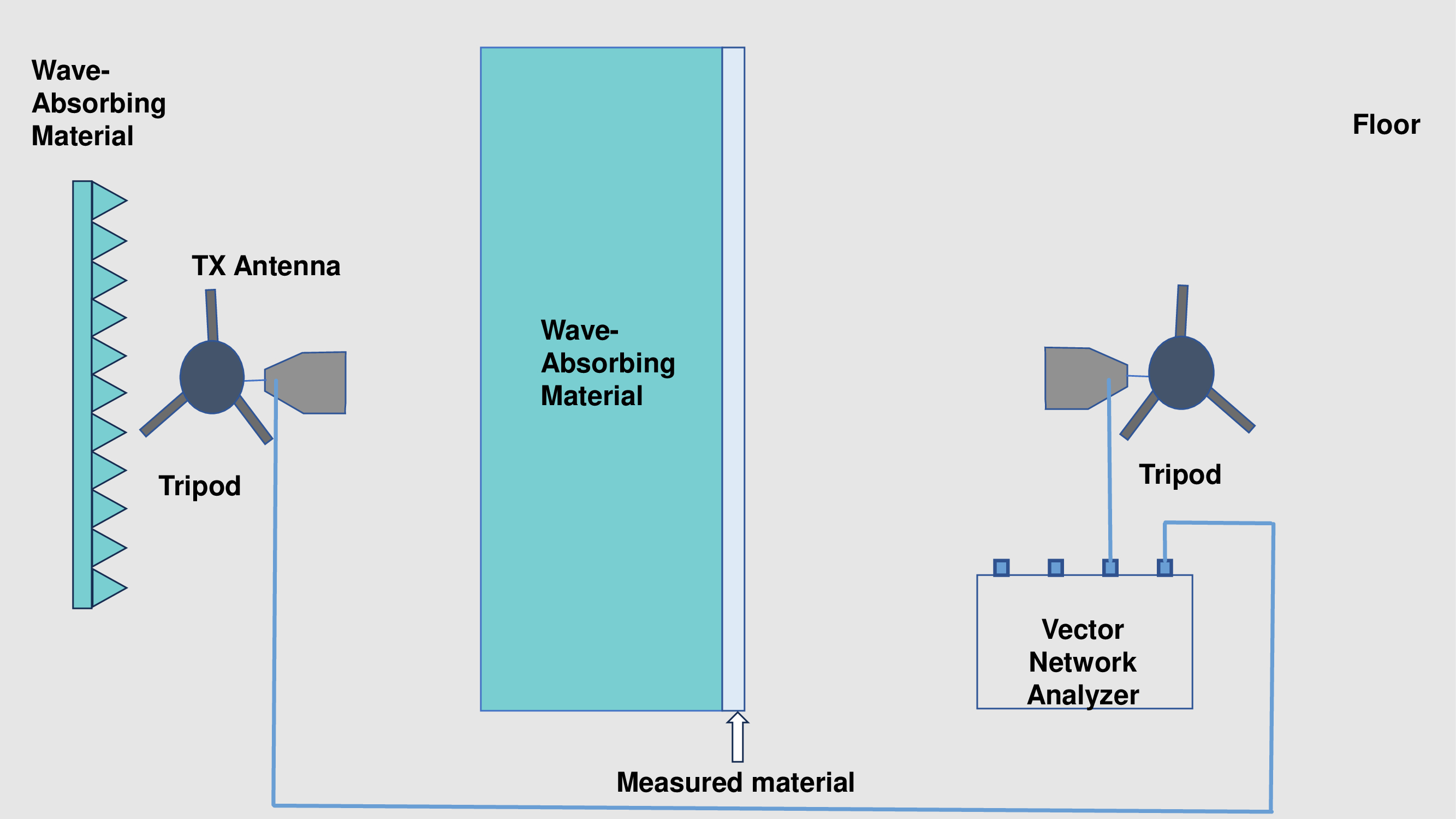}}
\caption{Schematic diagram of penetration loss measurement platform.}
\label{fig:1}
\end{figure}

The measurement platform is a frequency-domain channel measurement system based on a vector network analyzer (VNA) shown in Fig. \ref{fig:1}. The design of the platform is consistent with the green RF chain design direction of Integrated Sensing and Communications (ISAC) \cite{bib18}. This platform is composed of several modules and components, including a VNA, three pairs of antennas, synchronous cables, and wave-absorbing materials. Three pairs of antennas cover the 4–7 GHz, 7–10 GHz, and 10–16 GHz bands, respectively. This approach ensures comprehensive coverage across the specified frequency range. Pairs of antennas are connected using a sync cord and VNA. To minimize external interference and ensure accuracy, materials are placed at least 2.5 meters from the antennas, adhering to Rayleigh's distance formula. Measurement frequency starts at 4.5 GHz, cut-off frequency is 15.5 GHz, bandwidth is 1 GHz. To exclude the interference of indoor bypassing and reflecting and to make the signal from the antenna illuminate the measured materials as much as possible, the wave-absorbing materials are wrapped on the side of the measured material close to the transmitting antenna. The specific details of the measurement configuration for the measurement campaigns are shown in Table \ref{tabel:2}.
\begin{table}[ht]
\begin{center}
\caption{Measurement system parameters.}
\label{tabel:2}
\begin{tabular}{c|c}
\hline

\textbf{Parameters} & \textbf{Value/ Type}\\
\hline
Center Frequency & 4.5–15.5 GHz\\
Band Width & 1 GHz\\
TX/RX Antenna Type & Horn/ Horn\\
Antenna Height & 0.6 m\\
Distance From Antenna to Materials & 2.4 m\\
Antenna type&waveguide antenna\\
Horizontal and Vertical HPBW of Antenna 1 & $24^{\circ}$ and $24^{\circ}$\\
The Operating Frequency of Antenna 1&4.90–7.05 GHz\\
Gain of Antenna 1&15 dBi\\
XPD of Antenna 1&25 dB\\
Horizontal and Vertical HPBW of Antenna 2 & $26^{\circ}$ and $24^{\circ}$\\
The Operating Frequency of Antenna 2&7.05–10.0 GHz\\
Gain of Antenna 2&15 dBi\\
XPD of Antenna 2&35 dB\\
Horizontal and Vertical HPBW of Antenna 3 & $27^{\circ}$ and $25^{\circ}$\\
The Operating Frequency of Antenna 1&10.0–15.0 GHz\\
Gain of Antenna 3&15 dBi\\
XPD of Antenna 3&35 dB\\
\hline

\end{tabular}
\end{center}
\end{table}

\begin{table}[ht]
\begin{center}
\caption{Summary of Different Materials Settings.}
\label{tabel:1}
\begin{tabular}{|c|c|c|c|}
\hline

\multirow{2}{*}{\textbf{Materials}} & \textbf{Width} & \textbf{Height} & \textbf{Thickness}\\
 & (cm) & (cm) & (cm)\\
\hline
Wooden Board 1 & \multirow{2}{*}{130} & \multirow{2}{*}{130} & \multirow{2}{*}{1.0}\\
(pressed wooden board) &&&\\
\hline
Wooden Board 2 & \multirow{2}{*}{130} & \multirow{2}{*}{130} & \multirow{2}{*}{1.4}\\
(pressed wooden board) &&&\\
\hline
Wooden Board 3 & \multirow{2}{*}{130} & \multirow{2}{*}{130} & \multirow{2}{*}{1.0}\\
(solid wooden board) &&&\\
\hline
Glass 1 & \multirow{2}{*}{120} & \multirow{2}{*}{120} & \multirow{2}{*}{0.8}\\
(ordinary double-layer glass) &&&\\
\hline
Glass 2 & \multirow{2}{*}{120} & \multirow{2}{*}{120} & \multirow{2}{*}{0.8}\\
(frost glass) &&&\\
\hline
Foam Board 1 & \multirow{2}{*}{130} & \multirow{2}{*}{130} & \multirow{2}{*}{0.6}\\
(EVA foam board) &&&\\
\hline
Foam Board 2 & \multirow{2}{*}{130} & \multirow{2}{*}{130} & \multirow{2}{*}{0.8}\\
(EVA foam board) &&&\\
\hline
Foam Board 3 & \multirow{2}{*}{130} & \multirow{2}{*}{130} & \multirow{2}{*}{1.0}\\
(EVA foam board) &&&\\
\hline
Concrete Slab & \multirow{2}{*}{80} & \multirow{2}{*}{80} & \multirow{2}{*}{4}\\
(concrete) &&&\\
\hline

\end{tabular}
\end{center}
\end{table}

To investigate the penetration loss characteristics of different building materials in the FR1 and FR3 bands, ten typed building materials are selected, with some differing in type or size, and the specific parameter configuration is shown in Table \ref{tabel:1}. These materials were chosen because they are widely used in homes, offices and commercial buildings, so their penetration characteristics have a direct impact on the coverage performance of wireless communications in different building environments. Among them, No. 1-2 wooden boards are pressed wood boards, No. 3 is solid wood board, The different densities and structures of these boards result in different electromagnetic wave penetration losses. The No. 1 and No. 2 glass used in this measurement are common ordinary double-layer glass and frosted glass, respectively, their surface smoothness and multi-layer structure affect the transmission and reflection characteristics of electromagnetic waves. The EVA foam boards used in the measurement are common building foam materials.

\subsection{Measurement Methods and Procedures}
\begin{figure}[!ht]
\centering
\subfloat[The LOS scene.]{
		\includegraphics[scale=0.052]{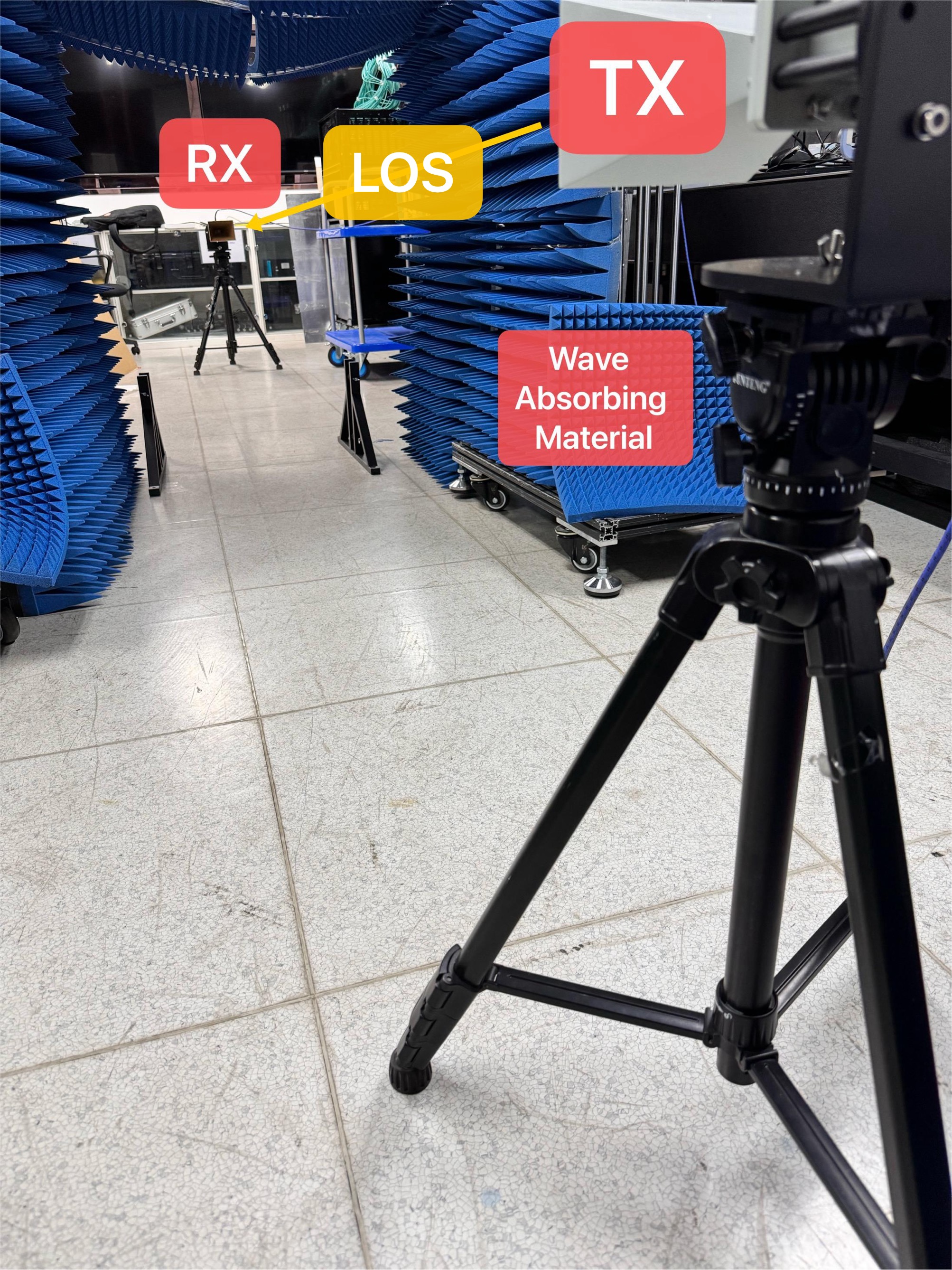}}
\hspace{0.2 cm}
\subfloat[The NLOS scene.]{
		\includegraphics[scale=0.0742]{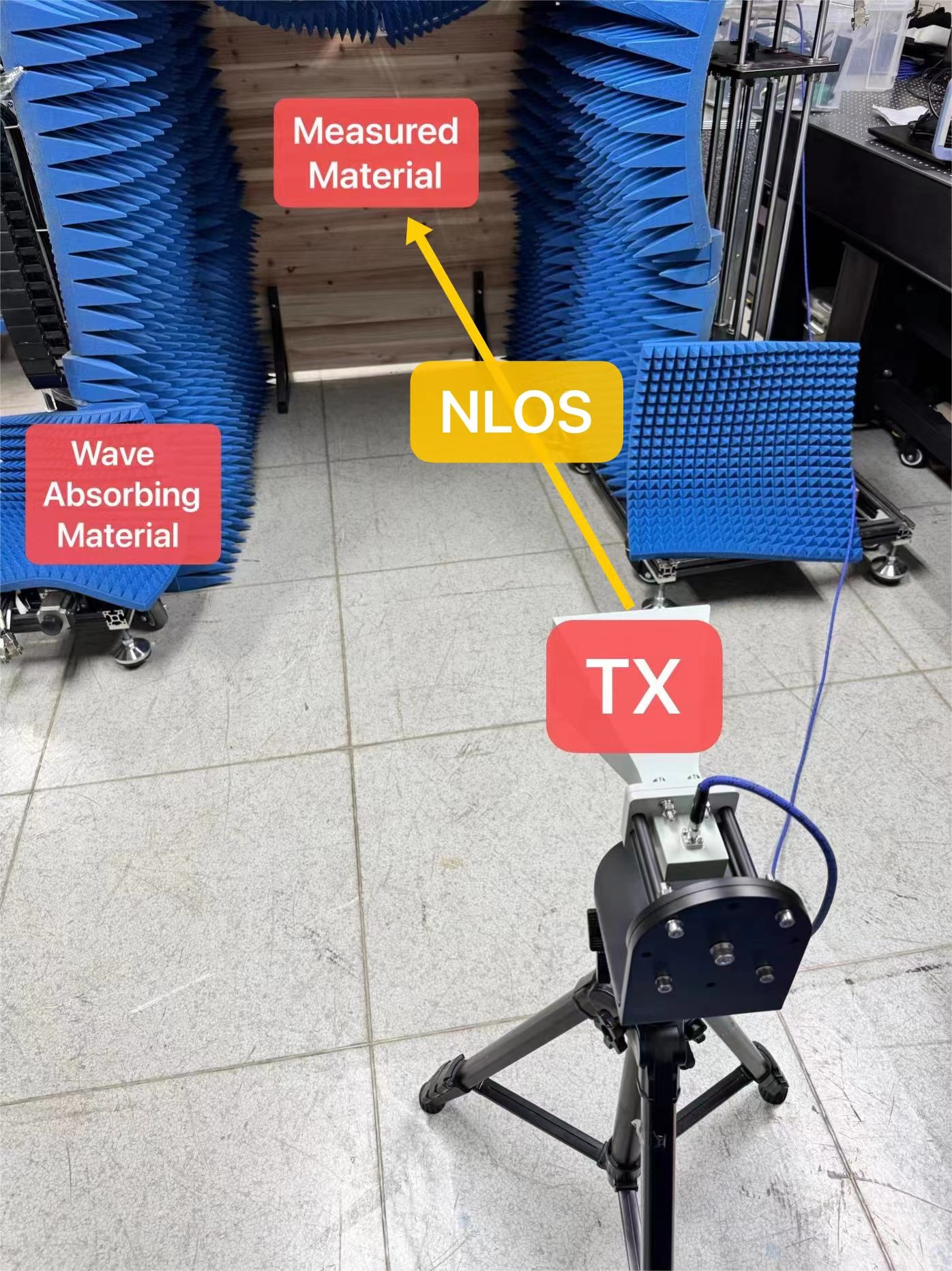}}
\caption{The penetration loss measurement scenario.}
\label{fig:2}
\end{figure}

The measurement environment is set up as shown below: place antennas operating in the corresponding frequency band at both sides of the material and connect them using a Vector Network Analyzer and synchronous cables\cite{bib2}. The Line-of-Sight (LOS) path is first measured without placing the materials, as shown in the left figure of Fig. \ref{fig:2}. Then, the measured material is placed between the two antennas, and the path loss is measured under Non-Line-of-Sight (NLOS) conditions, with the signal being obstructed by different materials, as shown in the right figure of Fig. \ref{fig:2}. 

We use a bandwidth of 1 GHz, the center frequency starts at 4.25 GHz and ends at 15.75 GHz in 1 GHz steps. For each broadband signal, we sample 256 points. We then apply the Fast Fourier Inverse Transform (IFFT) to the measured data to obtain the channel impulse response (CIR)\cite{bib4}. The CIR is given by:
\begin{eqnarray}
CIR(t) = \text{IFFT}\{D(f)\}, 
\label{formula:1}
\end{eqnarray}
where $D(f)$ represents the frequency-domain data sampled at 256 points. We select the path with the smallest delay as the measurement result, corresponding to the path directly penetrating the measured material. The penetration loss is obtained as the difference between the received power levels of the unobstructed path (LOS) and the received power level with the material obstructing the path between TX and RX (NLOS). We calculate the penetration loss using the S-parameters obtained for the LOS (line-of-sight) and NLOS (non-line-of-sight) conditions. The penetration loss is computed as:
\begin{eqnarray}
PL = S_{LOS} - S_{NLOS},
\end{eqnarray}
where $PL$ is measured penetration loss, $S_{LOS}$ and $S_{NLOS}$ are the S-parameters measured under LOS and NLOS conditions, respectively. This eliminates the effect of antenna gain on the results. Finally, we plot the change in penetration loss with frequency. The S-parameters obtained from each measurement are averaged from ten samples in the vector network analyzer.

\section{Penetration Loss Analysis and Modeling}
\subsection{Penetration Loss Analysis}

\begin{figure*}[!ht]
\centering
\subfloat[Wooden board.]{
		\includegraphics[scale=0.07]{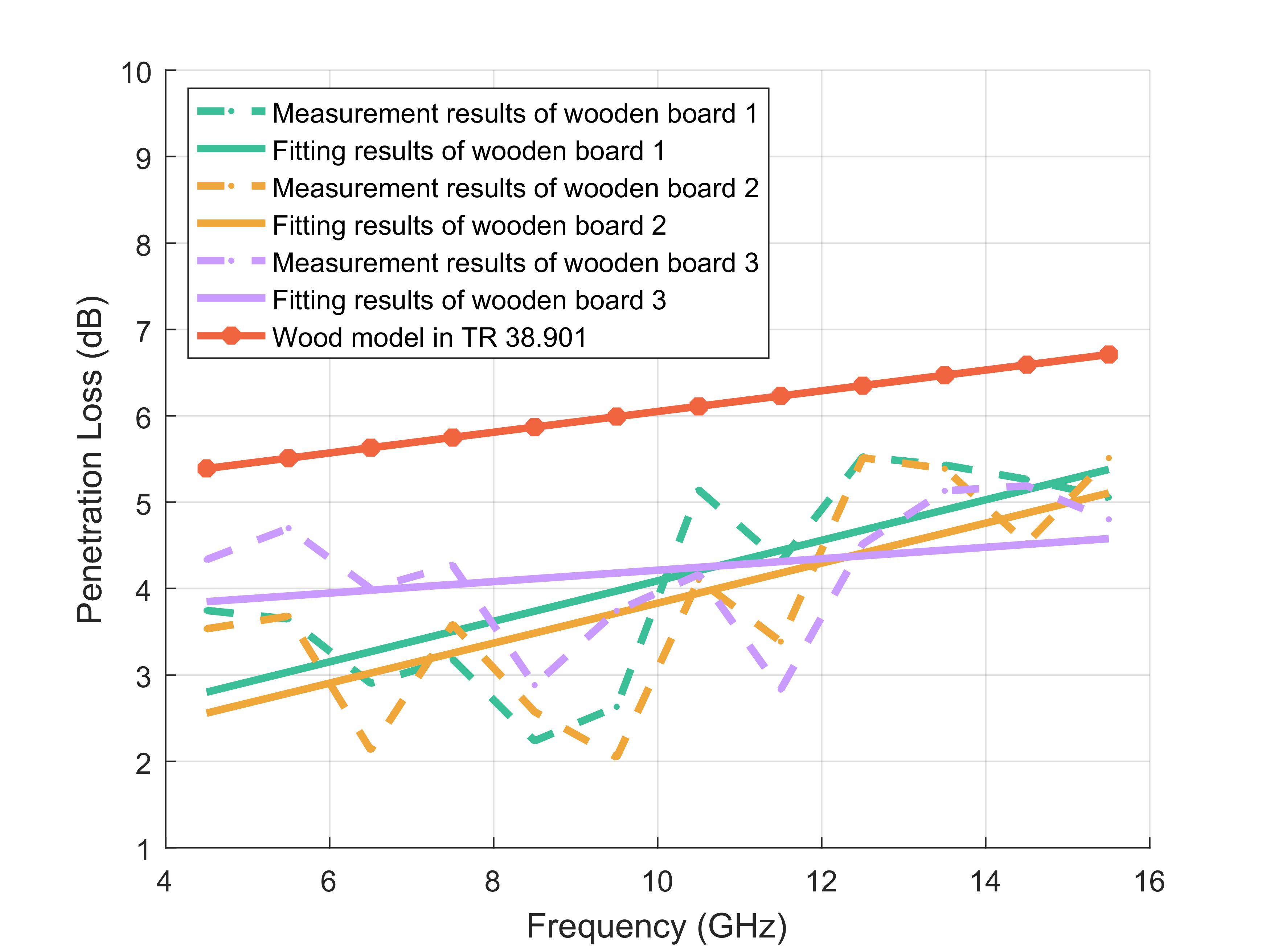}}
  \hspace{0.1cm} 
\subfloat[Glass.]{
		\includegraphics[scale=0.07]{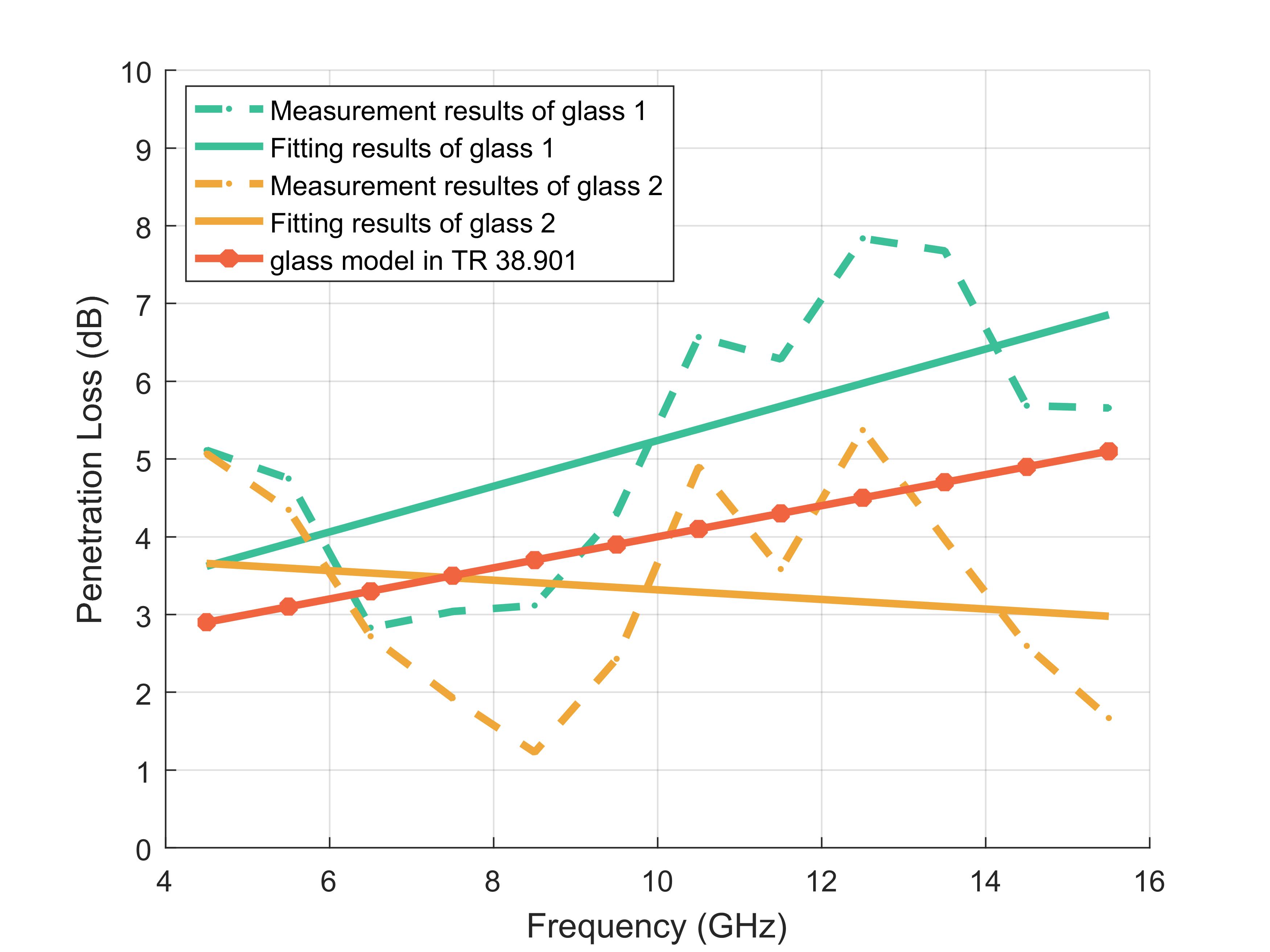}}
\\
\subfloat[Concrete slab.]{
		\includegraphics[scale=0.07]{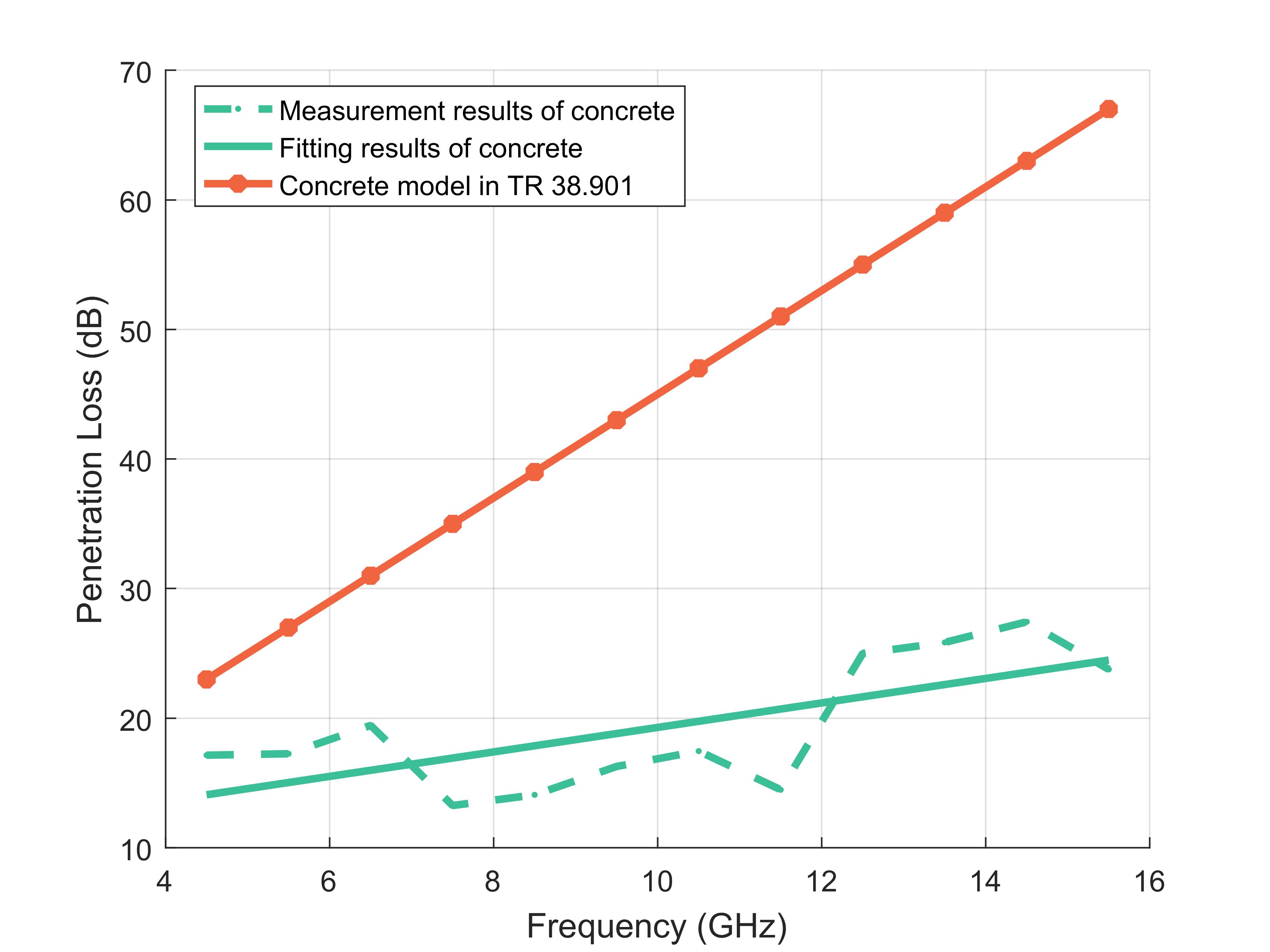}}
  \hspace{0.1cm} 
\subfloat[Foam board.]{
		\includegraphics[scale=0.07]{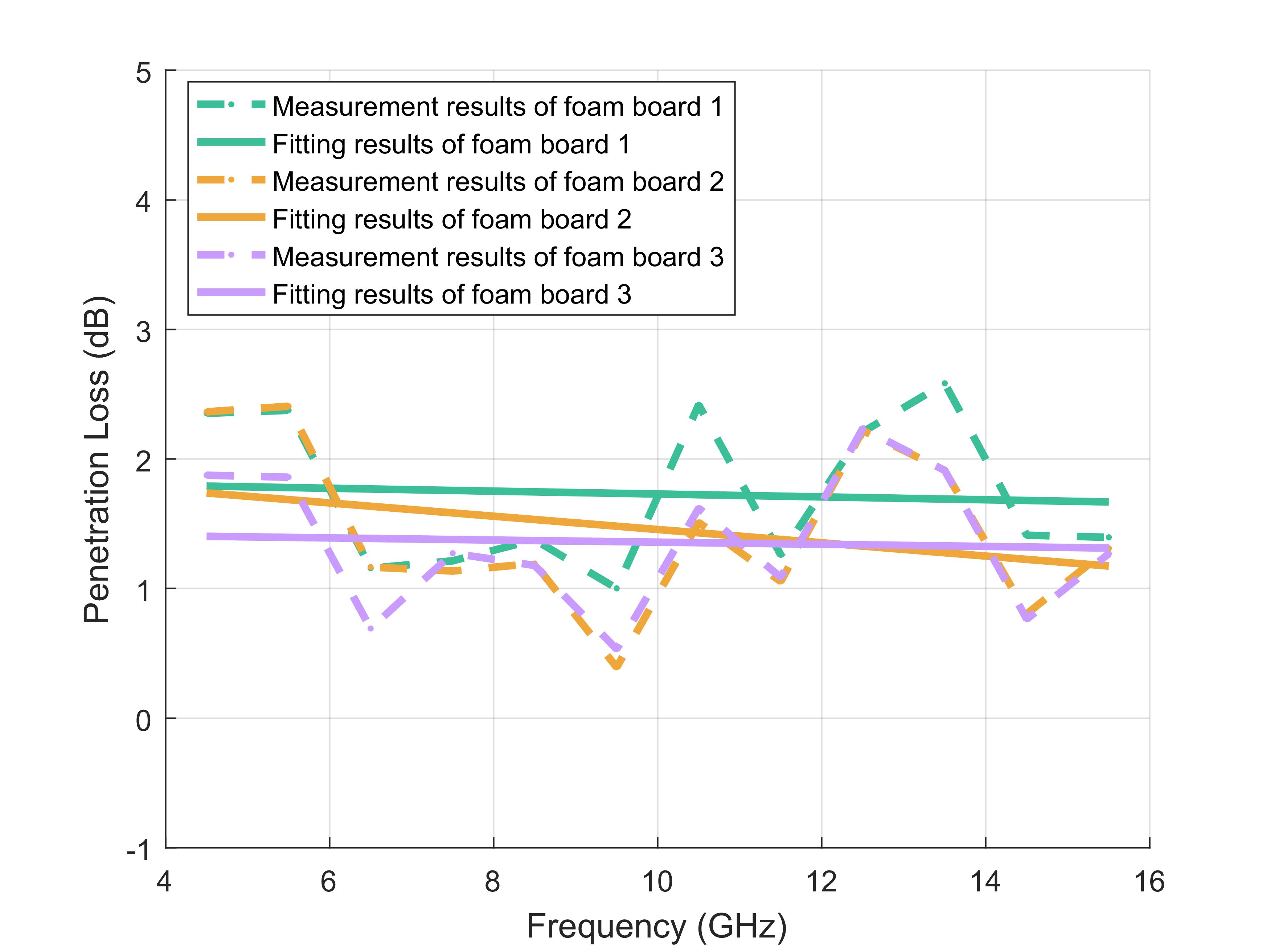}}
\caption{Measured Penetration loss and fitting results for different building materials.}
\label{Fig:3}
\end{figure*}

The data obtained from the measurements are processed to calculate the penetration loss values, which are then plotted. Additionally, the TR 38.901 standard is included for comparison, as shown in Fig. \ref{Fig:3}.

As shown in Fig. 3(a), the penetration loss values of the three different types of boards increase gradually with frequency. First, when comparing solid wood boards and pressed wood boards of the same thickness, the difference in penetration loss is minimal, within the range of 0 to 1.5 dB. Second, for boards made of the same material, the penetration loss increases with thickness, with a change range of approximately 2 to 5.5 dB. Finally, when these results are compared to the TR 38.901 model, the measured penetration losses are generally smaller by about 2 dB, this difference may be due to the use of different materials, resulting in similar trends but with slight variations in magnitude.

Fig. 3(b) illustrates the relationship between the penetration loss values of two types of glass. It is observed that both glasses exhibit an oscillating trend with frequency variations. There is a general decline in penetration loss in the frequency ranges of approximately 4–8 GHz and 12–16 GHz, while in the 8–12 GHz range, the penetration loss increases gradually with frequency. Overall, at the same frequency, ordinary double-layer glass has slightly greater penetration loss than frosted glass. The minimum values are reached at 6.5 GHz and 8.5 GHz, with values of 1.2 dB and 2.8 dB, respectively, while the maximum values are obtained both at 12.5 GHz, with values of 7.8 dB and 5.4 dB. The absolute difference in penetration loss between the two materials at the same frequency is approximately in the range of 0–2.4 dB, indicating that the penetration loss values are relatively close. Penetration loss of frosted glass tends to decrease with increasing frequency, probably due to the weakening of the scattering effect caused by surface roughness, the variation of the material dielectric constant with frequency, and the frequency dependence of the absorption effect. Comparison with the standard can be found, the value of the glass penetration loss does not completely obey the linear distribution. This is laterally confirmed by the simulation results in the literature \cite{bib12}, the reason for this phenomenon may be that there are cavities in the glass used for measurement, and the electromagnetic wave passes through materials with different dielectric properties when propagating.

Fig. 3(c) shows the penetration loss value of concrete versus frequency. It can be seen that the penetration loss value increases slowly with increasing frequency, ranging from about 13–27 dB. Compared with the TR 38.901 penetration model, it can be found that there are obvious differences between the two, and the difference in penetration loss at the frequency of 15.5 GHz is about 43 dB. The reason for this phenomenon may be that the penetration loss value of concrete in the standard is measured on the actual site, and the thickness is different. Not only that, although the trend for both is to increase with frequency, the rate of increase is quite different, possibly because the model is less applicable in this frequency band.

Fig. 3(d) depicts the relationship between penetration loss values and frequencies for three foam boards of different thicknesses. Comparative analysis shows that the penetration loss is relatively small. In addition, the penetration loss values of foam boards with different thicknesses do not completely conform to the law that the penetration loss increases with the increase in thickness, the penetration loss values tend to be stable with the increase of frequency, and the penetration loss values of foam boards with three thicknesses are not very different. This may be because foam boards usually have a lower dielectric constant, which means their ability to absorb high-frequency signals is weaker. Therefore, as the frequency increases, the loss of the signal when it penetrates the foam board may remain the same or even decrease.

Through comparative analysis, it is evident that the penetration loss increases with thickness. Except for the glass, the penetration loss of the materials measured in this study generally shows a linear relationship with frequency. Among wood, glass, foam, and concrete, foam has a weak obstructive effect on signals due to its structure, resulting in small loss when signals pass through it. Concrete, on the other hand, has the strongest obstructive effect on signals. Above all, a preliminary comparison of the model with the measured data, we can conclude that the extant 38.901 model does not predict penetration loss values in the FR1 and FR3 bands very well.

\subsection{Penetration Loss Modeling}
To provide a reference for the penetration loss values in the FR1 and FR3 frequency bands, we fit the measurements using a linear model, as expressed by the equation
\begin{eqnarray}
PL = k*f + b,
\label{formula:4}
\end{eqnarray}
where $PL$ represents the predicted penetration loss value, $f$ is the frequency, $k$ is the slope, indicating the change in $PL$ for each unit increase in f, and $b$ is the intercept. The specific details such as slope and intercept are shown in Table \ref{tab:3}. For wood and foam boards, the slopes are the same for different thicknesses of the same material, with a slight difference in intercepts (0.23 dB, 0.4 dB). 

\begin{table}[t]
    \centering
    \vspace*{0.1in}
    \caption{Comparison of fitting results for different materials with corresponding penetration loss models in TR 38.901.}
    \begin{tabular}{|c|c|c|c|}
        \hline
        \textbf{Category} & \textbf{Name} & \textbf{Slope (k)} & \textbf{Intercept (b)} \\
        \hline
        \multirow{4}{*}{Wood} & Wooden Board 1 & 0.23 & 1.75 \\
                              & Wooden Board 2 & 0.23 & 1.52 \\
                              & Wooden Board 3 & 0.07 & 3.55 \\
                              & TR 38.901 Wood Model & 0.12 & 4.85 \\
        \hline
        \multirow{3}{*}{Glass} & Double-Layer Glass & 0.30 & 2.30 \\
                                & Frosted Glass & -0.06 & 3.94 \\
                                & TR 38.901 Glass Model & 0.20 & 2 \\
        \hline
        \multirow{3}{*}{Foam} & Foam Board 1 & -0.01 & 1.84 \\
                               & Foam Board 2 & -0.05 & 1.97  \\
                               & Foam Board 3 & -0.01 & 1.44 \\
        \hline
        \multirow{2}{*}{Concrete} & Concrete Slab & 0.95 & 9.83 \\
                                   & TR 38.901 Concrete Model & 4.00 & 5.00 \\
        \hline       
    \end{tabular}
    \label{tab:3}
\end{table}

\begin{figure}[!ht]
    \centering
    \includegraphics[width=1\linewidth]{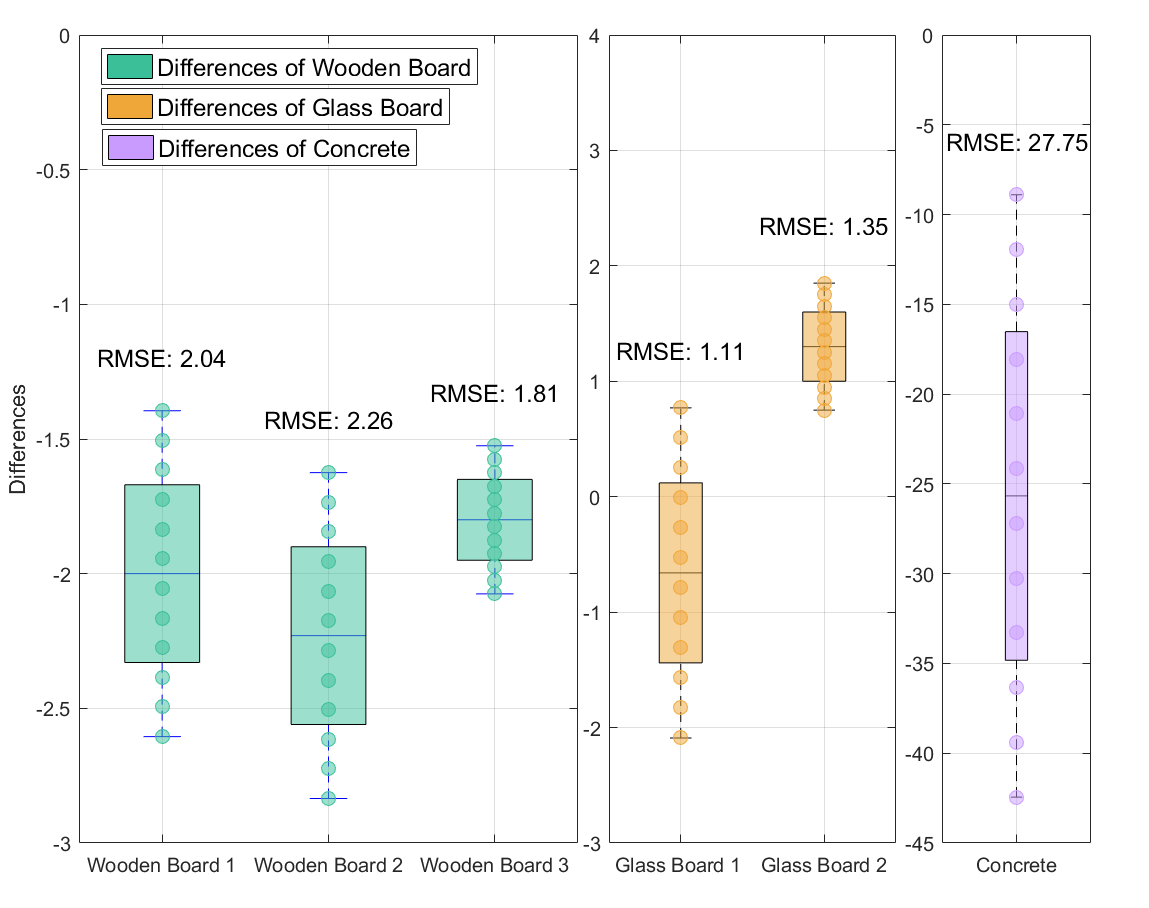}
    \caption{Difference and RMSE value distributions of penetration loss measurements between the fitted model and the standard model for different materials.}
    \label{fig:4}
\end{figure}

To assess the differences between the TR 38.901 model predictions and the measurement fitting results, we calculate the differences and RMSE using the fitting results of different building materials and the corresponding TR 38.901 model. The formulas are shown in equations (\ref{formula:5}) and (\ref{formula:6})
\begin{eqnarray}
e_{i} = y_{i} - \hat{y_{i}}, 
\label{formula:5}
\end{eqnarray}
\begin{eqnarray}
RMSE = \sqrt{\frac{\sum_{i=1}^{n} (e_{i})^{2} }{n}}, 
\label{formula:6}
\end{eqnarray}
where $y_{i}$ is the fitting value of the $i$ th measured penetration loss, and $\hat{y_{i}}$ is the value of the $i$ th predicted value (the value given by the TR 38.901 model), $n$ is the total number of observations. The calculation results are shown in Fig. \ref{fig:4}.

Fig. \ref{fig:4} shows the differences and RMSE between the fitted and standard wood, glass and concrete models based on dB-scale. For wood, the difference is centered around -2 dB, with minimal variation between boards of the same thickness (-2.7 to -1.35 dB). The RMSE values for wood ranged from 1.81 to 2.26 dB. This indicates high accuracy and low prediction error, with board 3 performing the best. Glass has a median difference close to 0 dB and low RMSE values (1.11 dB and 1.35 dB). Concrete has a wider spread of differences (-42.5 to -8 dB) and the largest RMSE (27.75 dB), indicating greater data dispersion for both.

By introducing the difference and the RMSE as evaluation indexes, we can find that the wood and glass standard model and the fitted model are not very different from each other. The two models for concrete have the most dispersed distribution of differences and the largest RMSE values. This indicates that the wood standard model is still applicable in the 4–16 GHz band. However, the concrete standard model needs to be modified. Additionally, since the glass model does not conform well to the linear model, it also needs to be revised to improve accuracy.

For wooden boards, the fitting results align closely with the penetration loss model for wood boards as outlined in TR 38.901. This consistency suggests that no significant modifications are necessary. However, the fitted model provides a more tailored approach for specific materials and thicknesses, offering valuable references when needed.

Regarding glass, the fitted results for frost glass and ordinary double-layer glass reveal distinct characteristics. Ordinary double-layer glass exhibits a slope of 0.30 and an intercept of 2.30, indicating a gradual increase of penetration loss with frequency compared to frost glass, which has a negative slope of 0.06 and a higher intercept of 3.94. The TR 38.901 glass model, with a slope of 0.2 and an intercept of 2, aligns more closely with ordinary double-layer glass but fails to accurately represent frost glass. Since ordinary double-layer glass is the most similar to the material used in the standard, the fitting model of ordinary double-layer glass is recommended for the FR1 and FR3 bands. However, the differences observed suggest that the linear model may not fully capture the behavior of the penetration loss characteristics of glass. Therefore, the nonlinear characteristics of glass will be explored in future research to improve the accuracy of the model.

In the case of concrete, the concrete fitting result shows a high slope of 0.95 and an intercept of 9.83, reflecting a steep increase in penetration loss with frequency. This is in stark contrast to the TR 38.901 concrete model, which has a much higher slope of 4 and a lower intercept of 5, indicating significant deviation from the measured data. This discrepancy underscores the inadequacy of the TR 38.901 model for accurately representing concrete's penetration loss. Consequently, the fitted model is recommended for using in this frequency range.

As for foam boards, the fitting results indicate slight variations in penetration loss characteristics among different samples. Due to these differences, establishing a uniform penetration loss model is currently challenging and would require extensive additional measurements. However, the fitted models can still serve as a useful reference for different foam board types and thicknesses.

In conclusion, the comparison of our fitted model with the 3GPP TR 38.901 standard model shows significant differences, especially for glass and concrete materials. These differences indicate that the TR 38.901 model cannot fully capture the actual penetration loss characteristics observed in the FR1 and FR3 bands.

\section{Conclusion}

In summary, this study addresses the limited understanding of penetration loss characteristics in the FR1 and FR3 bands. Measurements were carried out to assess the penetration loss of various building materials across these frequency ranges. Analysis reveals that penetration loss generally follows a linear distribution with frequency, except for glass, which exhibits a non-linear pattern characterized by an initial decrease, followed by an increase, and then another decrease. Furthermore, the existing TR 38.901 models are found to be inadequate in predicting penetration loss values within the FR1 and FR3 bands. To improve penetration loss prediction accuracy, we perform a linear fit on the measurements and derive new model parameters for different building materials.

\section*{Acknowledgment}
This work was supported in part by National Key Research and Development Program of China (2023YFB2904805), National Natural Science Foundation of China (62201086), Beijing Municipal Natural Fund (L243002), National Natural Science Foundation of China (62101069, 62341128), National Science Fund for Distinguished Young Scholars (61925102) and BUPT-CMCC Joint Innovation Center.

\vspace{12pt}

\end{document}